\newcommand{\uvby}{{\em uvby} }
\newcommand{\uvbyb}{{\em uvby}$\beta$ }
\newcommand{\hb}{H$\beta$ }
\newcommand{\ha}{H$\alpha$ }
\newcommand{\muno}{$m_{1}$ }
\newcommand{\cuno}{$c_{1}$ }
\newcommand{\co}{$c_{0}$ }
\newcommand{\byo}{$(b-y)_{0}$ } 
\newcommand{\mv}{$M_{V}$ }

\newcommand{\ecby}{$E^{\rm cs}(b-y)$ }
\newcommand{\eccu}{$E^{\rm cs}(c_{1})$ }
\newcommand{\db}{$\Delta \beta$}
\documentstyle[epsfig]{l-aa}     % LaTeX A&A  Version 3.0

\begin{document}

\thesaurus{  05 % A&A Section 5: Stellar clusters and associations
              (03.20.4;  % Techniques: photometric
              08.05.2;  % Stars: emission-line, Be
               10.15.1)  % open clusters and associations: general
   }
   \title{Be stars in open clusters.}
\subtitle{ III. A {\em uvby}$\beta$  calibration for the astrophysical
parameters of Be stars}

   \author{J. Fabregat \and J.M. Torrej\'{o}n}
 
   \offprints{J. Fabregat \\
(juan@pleione.matapl.uv.es) }
 
   \institute{Departamento de Astronom\'{\i}a, 
Universidad de Valencia, 46100 Burjassot, Valencia, Spain}

   \date{Received date; accepted date}
 
   \maketitle

   \begin{abstract}
We present linear relations between the equivalent widths of the Balmer
lines of Be stars and the anomalies in the \uvby photometric
indices produced by continuum circumstellar emission. A similar relation
exists when the emission in the \hb line is measured through the
photometric $\beta$ index.

These relations have been used to elaborate an empirical
calibration of the \uvbyb photometric system to determine the
intrinsic colours and indices, the relevant
astrophysical parameters and the absolute magnitude of the underlying B
star, valid for Be stars of spectral types earlier than B5. The
calibration  is based on
the study of 27 Be stars in 3 open clusters. The proposed calibration
procedure allows the determination of the interstellar
reddening with an accuracy of 0.033 mag. (rms) and the absolute magnitude
with an accuracy of 0.7 mag. 

Our calibration independently confirms the previously known result that Be
stars of spectral
types in the range B0-B5 are overluminous by a mean of 0.3 mag. with
respect to their absorption-line counterparts of the same spectral type.

   \end{abstract}

\section{Introduction}

The \uvby and \hb photometric systems, defined by Str\"{o}mgren (1966) and
Crawford \& Mander (1966), were designed to measure fundamental spectral
signatures in early- and intermediate-type stars. Their combined use is
one of the most suitable photometric methods to determine temperatures,
luminosity classes, absolute magnitudes and other stellar parameters of
astrophysical interest. 

In the case of Be stars, however, the usual techniques of \uvbyb
calibration are not suitable. In their observed photometric indices,
besides the contribution of photospheric emission and interstellar
extinction, there is an additional emission contribution from a
circumstellar envelope. 

The main purpose of this work is to present a method to 
determine such intrinsic parameters from \uvbyb photometry. It has been
shown by several authors (Dachs et al. 1986, 
1988; Kaiser 1989) that the continuum emission of the circumstellar envelope 
is closely correlated with 
the equivalent width of the Balmer lines. Thus, by measuring the Balmer 
emission-line strengths, the underlying star 
contribution to the photometric indices can be decoupled from 
the circumstellar disk contribution, and then the usual \uvbyb 
calibrations can be applied. A preliminary exposition of this method is
given by Fabregat \& Reglero (1990), hereafter referred to as FR90.

In order to accurately determine the relationship between the circumstellar 
continuum emission and the emission in the Balmer lines, we have
developed an 
observational programme of simultaneous \uvbyb photometry and Balmer line  
spectroscopy of Be stars in open clusters. The advantage of studying stars
in clusters is that their intrinsic colours and distances can be
inferred from the cluster parameters, and thus the anomalies in their
photometric indices caused by the continuum emission from the envelope can
be directly measured. The \uvbyb photometry is presented in Fabregat et
al. (1996), hereafter
referred  to as Paper I. The Balmer line spectroscopy is presented in
Torrej\'{o}n et al. (1997), hereinafter Paper II. 

\section{Relationship between photometric anomalies and Balmer line
equivalent widths}

In Paper I we showed that Be stars occupy anomalous positions in the 
photometric diagrams, which can be explained in terms of 
the circumstellar continuum radiation contribution to the photometric 
indices. 
In the \mv -- \byo plane Be stars appear redder than the non emission B
stars, 
due to the additional reddening caused by the hydrogen free-bound and 
free-free recombination in the circumstellar envelope. In the \mv -- \co 
plane 
the earlier Be stars present lower \co values than absorption-line B stars, 
which 
is caused by emission in the Balmer discontinuity, while the later Be stars 
deviate towards higher \co values, indicating absorption in the 
Balmer discontinuity of circumstellar origin.

Following FR90, we will denote \ecby and \eccu the contribution of the
circumstellar emission to the $(b-y)$ colour and \cuno index. The values 
of
these circumstellar excesses can be directly measured for Be stars in open
clusters. We consider the circumstellar excess value as the difference
between the observed photometric index and the photometric index
corresponding to the absorption-line B stars of the same absolute
magnitude. The loci of the absorption-line B stars in the photometric 
\mv -- \byo and \mv -- \co diagrams has been considered to be well
represented by the cluster isochrones. In Figure 1 we show graphically how
the \ecby and \eccu values are measured. The main source of error in the
determination of the circumstellar excesses, in addition to the errors of
the photometric measurements and the cluster reddening, is the assumption
of the isochrone as being representative of the normal star positions. To
estimate this error, we have computed the standard deviation of the
differences between the positions of the absorption-line B stars and the
isochrone,
measured as explained in Figure 1. The obtained values are 0.025 and 0.033
mag. for $(b-y)$ and \cuno respectively. We assume these values as the
errors of our \ecby and \eccu determinations.

%------------------------------------------------------------------------
\begin{figure}
\begin{center}
\leavevmode
\epsfig{file=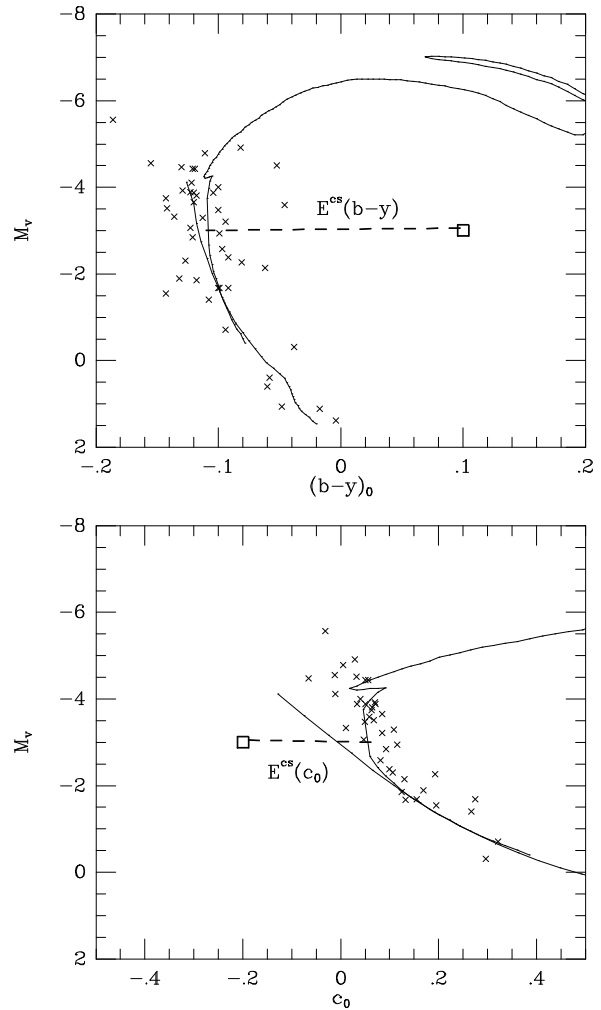, width=8cm,  bbllx=40pt,  bblly=290pt,
    bburx=300pt,  bbury=740pt,  clip=}
\end{center}
\caption{Graphical interpretation of the \ecby and \eccu photometric
excesses for a given young open cluster. Lines represent the ZAMS and
cluster isochrone, crosses the absorption-line stars and the open square a
given Be star for which the photometric excesses are being measured. 
The cluster represented is $h$ \& $\chi$ Persei, and the
photometric data are from Paper I.}
\end{figure}
%------------------------------------------------------------------------
  
It should be noted that with this approach we assume
that there are not significant variations of the absolute
magnitude of circumstellar origin. To check this assumption, in Figure 2
we have plotted the excess in \mv for the Be stars in our sample
against the \ha equivalent width. The \mv excess has been computed as the
difference between the absolute magnitude derived from the
observations and using the cluster reddening and distance modulus, and the
mean \mv for the spectral type of each star. We have assumed the spectral
types determined by Slettebak (1985). As it can be seen, no trend of
absolute magnitude variation with the amount of circumstellar emission is
apparent. On the other hand, Zorec \& Briot (1991) have studied in depth
the overluminosity effects produced by the presence of the circumstellar
emission. They concluded that the mean \mv variation amounts between --0.5
and --0.1 mag., for spectral types between B0 and B5. As it can be seen in
Figure 1, this variation would introduce very small errors in the
determination of \ecby and $E^{\rm cs}(c_{1})$, well within the errors
estimated in the above paragraph.

%------------------------------------------------------------------------
\begin{figure}
\begin{center}
\leavevmode
\epsfig{file=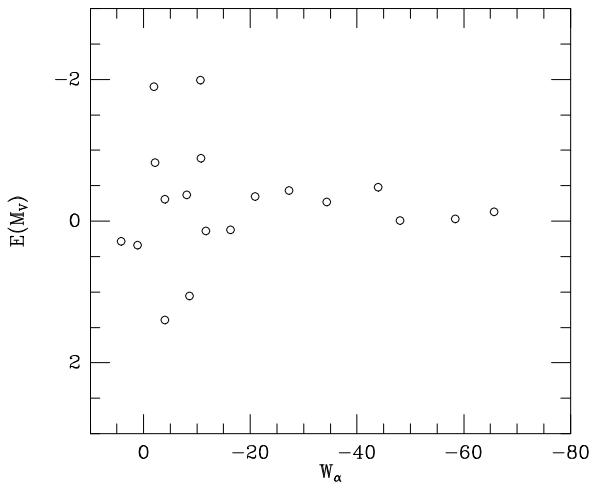, width=8cm,  bbllx=40pt,  bblly=510pt,
    bburx=305pt,  bbury=740pt,  clip=}
\end{center}
\caption{Excess in the absolute visual magnitude versus the emission line 
strength.}
\end{figure}
%------------------------------------------------------------------------

With the above precepts we have measured
the \ecby and \eccu values for all Be stars in the clusters $h$ and
$\chi$ Persei, NGC 663, Pleiades, NGC 2422 and $\alpha$ Persei for which 
\uvbyb photometry is given in Paper I. We have excluded stars 107 and
110 in NGC 663, because their peculiar position in the photometric
diagrams makes the determination of the excesses very uncertain. We have
also considered the cluster isochrones computed in Paper I.

The amount of circumstellar continuum emission is closely correlated with
the equivalent widths of the Balmer emission lines (Dachs et al. 1986,
1988; Kaiser 1989). Thus, the circumstellar excesses in the photometric
indices should be also correlated with the equivalent widths. We have
determined these relations by comparing our values of \ecby and \eccu with
the \ha and \hb equivalent widths presented in Paper II. In Paper I we
have shown that Be stars earlier and later than B5 present different
behavior in the photometric diagrams, showing \eccu of
opposite sign. In consequence, we have considered separately early and
late Be stars.

For the late Be stars no clear relations are present between the
photometric excesses and the equivalent widths. This may be caused by the
small number of stars considered, only eight. Therefore we will restrict
our study to Be stars earlier than B5, which implies that we will
consider only stars in the clusters $h$ and $\chi$ Persei and NGC 663.
From this sample we obtain the following relations:

\begin{eqnarray}
\nonumber & \hspace{5mm}\sigma & \hspace{3mm}r \\ 
E^{\rm cs}(b-y) = -0.0014 W_{\rm e}({\rm H}\alpha) + 0.014  &
\hspace{5mm}0.043 & 0.61\\
E^{\rm cs}(c_{1}) = 0.0019 W_{\rm e}({\rm H}\alpha) - 0.031 &
\hspace{5mm}0.038 & 0.77\\
E^{\rm cs}(b-y) = -0.0086 W_{\rm e}({\rm H}\beta) + 0.042   &
\hspace{5mm}0.045 & 0.54\\
E^{\rm cs}(c_{1}) = 0.0200 W_{\rm e}({\rm H}\beta) - 0.057  &
\hspace{5mm}0.030 & 0.91
\end{eqnarray}

where the equivalent widths, expressed in \AA , are considered
negative for emission lines. $\sigma$
represents the standard deviation of the relation and $r$ the correlation
coefficient. These relations have been obtained with equivalent widths not
corrected
for photospheric absorption, as presented in Paper II. We have also
computed the above relations for emission equivalent widths. To correct
for photospheric absorption we have performed the following iterative
procedure:

\begin{itemize}

\item We have used the above relations to compute
the circumstellar excesses for each star, and corrected the
photometric indices for circumstellar effects. 

\item With the corrected indices we have computed the $T_{\rm eff}$ and
$\log g$ for each star, using the formulae given by Balona
(1994) to interpolate in the grid of theoretical \uvby indices
computed  by Lester et al. (1986). 

\item With these values we have
computed the equivalent widths of the photospheric absorption lines by
means of the Kurucz (1979) atmosphere models. The obtained values have
been subtracted from the uncorrected equivalent widths. Doing this, we are
adding to the measured emission line above the continuum the amount of
emission filling-in the photospheric absorption line. 

\item We have computed new correlations, and repeated all the procedure
until the result converges.

\end{itemize}

The final relations for corrected equivalent widths are the following:

\begin{eqnarray}
   & \hspace{5mm}\sigma & \hspace{3mm}r \nonumber \\
E^{\rm cs}(b-y) =  -0.0015 W_{\rm e}({\rm H}\alpha) + 0.009  &
\hspace{5mm}0.043 & 0.63 \\
E^{\rm cs}(c_{1}) =  0.0019 W_{\rm e}({\rm H}\alpha) - 0.025 &
\hspace{5mm}0.038 & 0.76 \\
E^{\rm cs}(b-y) =  -0.0085 W_{\rm e}({\rm H}\beta) + 0.019   &
\hspace{5mm}0.045 & 0.53 \\
E^{\rm cs}(c_{1}) =  0.0199 W_{\rm e}({\rm H}\beta) - 0.025  &
\hspace{5mm}0.029 & 0.92
\end{eqnarray}
 
The slopes of the corrected relations remain practically unchanged with
respect to the uncorrected ones. The only significant differences are in
the zero points, which in the corrected equations are compatible with
zero within the errors, as expected because in this last case we are
comparing two effects produced by the presence of a circumstellar
envelope. These relations are presented in Figure 3. Equations (1) and (5)
closely correspond to Eq. (39) in Dachs et al. (1988), which represents 
the relationship between the circmstellar excess in the $(B-V)$ colour and
the \ha equivalent width, found by these authors.

%------------------------------------------------------------------------
\begin{figure*}
\begin{center}
\leavevmode
\epsfig{file=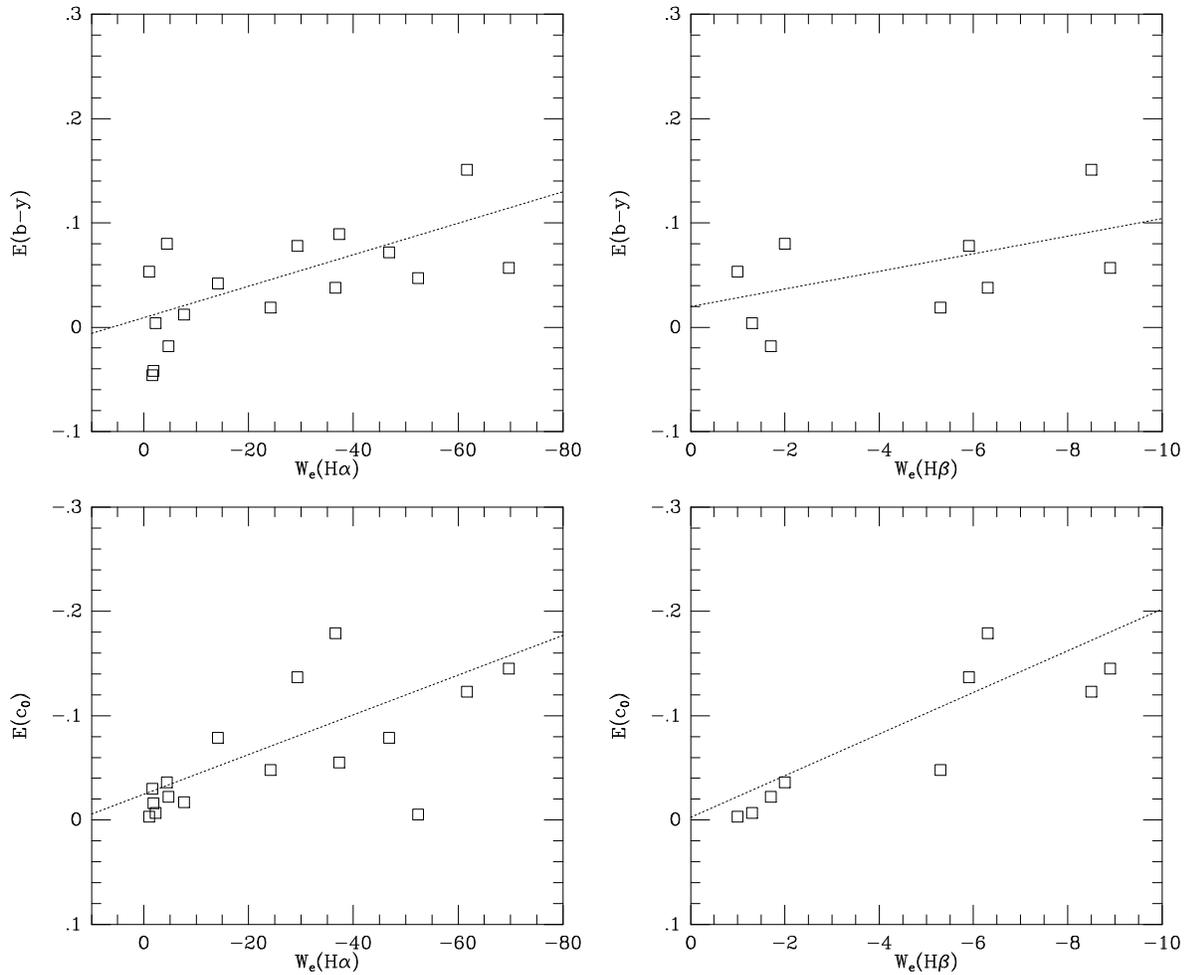, width=16cm,  bbllx=40pt,  bblly=300pt,
    bburx=580pt,  bbury=750pt,  clip=}
\end{center}
\caption{Relationship between the photometric circumstellar excesses and
the emission line equivalent widths corrected for photospheric
absorption.}
\end{figure*}
%------------------------------------------------------------------------

From the $\sigma$ and $r$ coefficients of the above relations and from
Figure 3, the correlation between the photometric excesses and the
equivalent widths is apparent. The correlation seems to be better for the
\hb equivalent width. This would mean that the region of the circumstellar
disc in which the continuum emission contaminating the \uvby indices
arises is closer to the region in which the \hb emission line is formed.
This is expected from the fact that the wavelength of the \hb line is
closer to the \uvby bandpasses.

If the \hb equivalent width is closely correlated with the circumstellar
continuum emission, the photometric $\beta$ index has to be correlated as
well. The $\beta$ index is, by construction, linearly related to the \hb
equivalent width (Golay, 1974). Furthermore, Fabregat \& Torrej\'{o}n
(1997) have
shown that this linear relation is valid even for emission-line stars.
Then we have also studied the correlation between the $\beta$ index and
the 
excesses in the photometric indices. As in the case of the spectroscopic
equivalent widths, the value to be correlated with \ecby and \eccu is not
the $\beta$ index, but the excess in the $\beta$ index due to the
circumstellar emission, which we will denote \db. {\db } is defined as the
difference between the observed $\beta$ index and the $\beta$ index of the
underlying B star. The latter is not known {\em a priori}, and we
have to estimate its value by means of an iterative procedure. As this
discussion applies to a rather narrow range of spectral types, O9 to B5,
we have considered a mean value for the $\beta$ index in this spectral
range, namely $\beta$ = 2.63. We obtain a first approximation to the
$\beta$ excess as {\db } = $\beta$ -- 2.63, where $\beta$ is the observed
value. Then we correlate these values with the photometric excesses, and
obtain a first approximation to \ecby and \eccu. With the latter value
we obtain \co corrected from circumstellar effects, and from it we
determine the $\beta$ value of the underlying star -which we will denote 
$\beta_{*}$ in order to avoid confusion with the observed $\beta$ value
-by means of the following relation

\begin{equation}
\beta_{*} = 2.620 + 0.2517 c_{0} - 0.1400 c_{0}^{2} + 0.1704 c_{0}^{3}
\end{equation}

(Balona \& Shobbrook, 1984). The $\beta_{*}$ values obtained with this
equation can safely replace the observed $\beta$ indices in the range of
the   early B-type main sequence and giant stars (Balona 1994). With the
$\beta_{*}$ values obtained in this way we re-determine \db = $\beta -
\beta_{*}$, and repeat the procedure until convergence.

As the relations we are now studying are only between photometric indices,
we have searched the literature for \uvbyb photometry of Be stars in
open clusters, in order to derive the relations from as large a sample as
possible. We have used, as well as our photometry for $h$ and $\chi$
Persei and  NGC 663 (Paper I), data for Be stars in the clusters NGC 3766
(Shobbrook
1985, 1987) and NGC 4755 (Perry et al. 1976).

%------------------------------------------------------------------------
\begin{figure*}
\begin{center}
\leavevmode
\epsfig{file=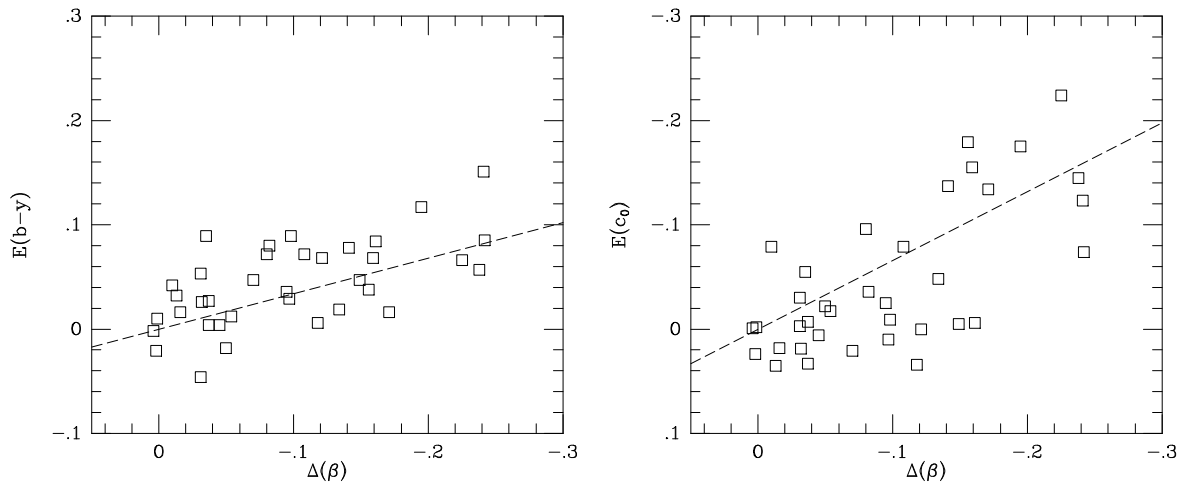, width=16cm,  bbllx=40pt,  bblly=500pt,
    bburx=580pt,  bbury=750pt,  clip=}
\end{center}
\caption{Relationship between the \uvby circumstellar excesses and
the excess in the $\beta$ index.}
\end{figure*}
%------------------------------------------------------------------------

In the relation obtained after the iterative procedure the zero points are
again compatible with zero. Then, with the final {\db } values we have
computed the correlations assuming no zero point. The results are the
following:

\begin{eqnarray}
\nonumber & \hspace{30mm}\sigma \\
E^{\rm cs}(b-y) = -0.339  \Delta \beta & \hspace{30mm}0.033 \\
E^{\rm cs}(c_{1}) = 0.661 \Delta \beta & \hspace{30mm}0.051 \\
E^{\rm cs}(m_{1}) = 0.125 \Delta \beta & \hspace{30mm}0.024 
\end{eqnarray}

The two first relations are presented in Figure 4. For the sake of
completeness, we have also computed the relation for the \muno index.

Balona \& Shobbrook (1984) proposed an alternative to Equation (9) to
determine $\beta_{*}$ for stars of luminosity class IV-III. We have
reproduced all the above procedure and the discussion in Section 4 using
different equations for stars of luminosity class V or IV-III. The results 
obtained are basically the same that those presented here (Torrej\'on
1997).
We have preferred to construct our calibration using only Equation (9),
because in this way we do not need a previous knowledge of the star's
luminosity class.

\section{The \uvbyb calibration}

In the previous section we have derived relations between the emission in
the \hb line, measured through the $\beta$ index, and the 
excesses in the photometric \uvby indices of circumstellar origin. The
existence of such relations
opens the possibility to use the \hb photometry to correct the \uvby
indices for circumstellar emission, and then derive the
interstellar reddening and the intrinsic indices of the
underlying star. These intrinsic indices can be used to estimate the
astrophysical parameters of the underlying star by means of the usual
\uvby calibrations. 

\subsection{Interstellar reddening and intrinsic colours}

The procedure to obtain the intrinsic \uvby indices of a B star,
correcting for the effect of the interstellar reddening, has been given by
Crawford (1978). In the case of a Be star, we have to correct for both
interstellar reddening and circumstellar continuum emission. In
this section we propose a procedure to determine the
circumstellar excesses, the interstellar reddening and the intrinsic
colours of the underlying star. The basic idea is to use
the $\beta$ index as additional information to characterize the
circumstellar emission, and then use the Crawford (1978) method to measure
the interstellar reddening. As both effects are coupled, there is not an
easy way to decouple the effects of circumstellar and interstellar
reddening. We propose an iterative procedure to determine both of them,
which consist in the following steps:

\begin{itemize}

\item Make a first aproximation of {\db } = $\beta$ -- 2.63. Substitute
this value in Equations (10) and (11) to obtain \ecby and \eccu. 
Whit this
values obtain $(b-y)$ and \cuno corrected for circumstellar excess.

\item Whit the above $(b-y)$ and \cuno obtain the interstellar reddening
and the intrinsic colours and indices, by means of the Crawford (1978)
procedure.

\item Use the intrinsic \co index to estimate the value of $\beta_{*}$
by means of equation (9). 

\item Make a new determination of \db = $\beta - \beta_{*}$. Repeat all
the procedure until convergence.

\end{itemize}

\subsection{Luminosity calibration}

The $\beta$ index is the main luminosity indicator for OB stars in the
\uvbyb photometric system. In the case of Be stars, however, the $\beta$
index is contaminated by the circumstellar emission, and is not useful to
estimate the luminosity. Moreover, in the calibration method we are
proposing, the $\beta$ index has been used to characterize the
circumstellar emission and correct the other indices for circumstellar
effects. There is not any other independent index which can be related
with
the stellar luminosity, and therefore there is not any way to directly
determine the luminosity within our approach.

It has to be considered, however, that the present calibration is to be
applied to a rather narrow range of luminosity classes. When studying
absorption-line OB stars, there is a wide range of variation of the
$\beta$ index for a given $c_{0}$, originated by the differences in the
\hb
line strenght between supergiants and main sequence stars. Our calibration
deals with Be stars, which are, by definition, restricted to luminosity
classes III to V. As we stated in Section 2, within this range of
luminosities
the value of $\beta$ can be inferred from \co by means of Equation (9).
Then we propose to estimate the absolute magnitude of Be stars
from the \co index alone, by means of the following procedure:

\begin{itemize} 

\item Derive the $\beta_{*}$ value from \co by means of Equation (9).

\item Substitute the values of \co and $\beta_{*}$ in the Balona \&
Shobbrook (1984) calibration of absolute magnitude. 

\end{itemize}
 
\section{Discussion}

To estimate the accuracy of the above procedures we have computed the
intrinsic indices, interstellar reddening and absolute magnitude for all
the cluster Be stars with available \uvbyb photometry used in 
Section 2, by means of the method described in Section 3. 
We have compared
the obtained reddening values with the mean reddening of each
cluster, and the obtained \mv with the value directly derived from the
photometry and the cluster reddening and distance. We have used the same
values for the clusters' reddening and distance modulus as in Paper I.
References for these values are also given in Paper I. The results are
shown in Table 1. 
In Table 2 we compare the mean values for each cluster obtained with our
procedure and the values from the literature given in Paper I. Columns 4
and 5 are the mean values of the reddening and distance modulus obtained
with the procedure presented in Section 3 for all the Be stars in each 
cluster with available \uvbyb photometry. Columns 6 and 7 are the same
for reddening and distance modulus values obtained with the FR90
calibration. Note that the $DM$ values have been computed from the 
absolute magnitudes in the usual way, without taking into account the 0.3
mag. correction suggested below in Eq. (14). 

%-----------------------------------------------------------------------
\begin{table}
\begin{center}
\caption{Comparison between the $E(b-y)$ and \mv values obtained with our 
calibration and the values derived from the clusters' reddening and
distance.} 
\begin{tabular}{rrrrrr}
\hline
  & & & & & \\ 
Star & $\beta_{*}$ & $E(b-y)$ & \small{$\Delta E(b-y)$} & 
$M_{V}(\beta_{*})$ & \small{$\Delta(M_{V})$}\\
  & & & & & \\ 
\hline
    &   &   &  &  &\\
\multicolumn{6}{c}{$h$ \& $\chi$ Per} \\
  & & & & & \\ 
     49  &      2.619 &   0.362 &  -0.048 &  -3.02 & 0.77\\
    309 &      2.604 &   0.396 &  -0.014 & -3.65  & -0.15\\
    717 &      2.625 &   0.366 &  -0.044 &  -2.80 &   0.84\\
   847 &      2.634 &   0.421 &   0.011 &  -2.50 &   1.30\\
    1161&      2.632 &   0.454 &   0.044 &  -2.55 &   0.18\\
    1261&      2.624 &   0.463 &   0.053 &  -2.84 &   0.49\\
    1268&      2.622 &   0.444 &   0.034 &  -2.89 &   0.64\\
    1702&      2.631 &   0.372 &  -0.038 &  -2.60 &   0.61\\
    1926&      2.608 &   0.460 &   0.050 &  -3.44 &  -0.45\\
   2138&      2.622 &   0.454 &   0.044 & -2.90  &  0.73\\
   2165&      2.610 &   0.439 &   0.029 & -3.36  & -0.43\\
   2284&      2.625 &   0.377 &  -0.033 &  -2.80 &    0.42\\
    2262&      2.636 &   0.427 &   0.017 &  -2.44 &  -0.12\\
     2371&      2.625 &   0.392 &  -0.018 &  -2.80 &   0.90\\
     2563&      2.616 &   0.422 &   0.012 &  -3.12 &  -0.75\\
     2566&      2.645 &   0.423 &   0.013 &  -2.15 &   0.25\\
     2649&      2.648 &   0.358 &  -0.052 &  -2.07 &   0.50\\
     2804&      2.615 &   0.453 &   0.043 &  -3.19 &  -0.72\\
  & & & & & \\
\multicolumn{6}{c}{NGC 663} \\
  & & & & & \\ 
    2 &       2.636&    0.611&    0.020 &  -2.42 &   0.07\\
    6 &       2.629&    0.657&    0.067&   -2.65&    0.04\\
    10&       2.622&    0.624&    0.034&   -2.90&   -0.05\\
     21&       2.643&    0.566&  -0.024&   -2.21&    1.95\\
     93&       2.657&    0.559&   -0.031&   -1.83&    0.94\\
    &  &   &   &  \\
\multicolumn{6}{c}{NGC 3766} \\
  & & & & & \\ 
           1&    2.661&    0.116 &  -0.034  & -1.74  &  1.71\\
           15&    2.672 &   0.117 &  -0.033 &  -1.50  &  1.99\\
           26 &   2.665 &   0.140  & -0.010 &  -1.66  &  1.21\\
           27 &   2.654&   0.132 &  -0.018 &  -1.92&    1.67\\
            63&    2.659  &  0.139 &  -0.011 &  -1.79  &  1.01\\
            81&    2.672 &   0.133 &  -0.017 &  -1.50  &  0.48\\
           239 &   2.653 &   0.141& -0.009 &  -1.93 &   0.67\\
  &  &  &   &   &  \\
\multicolumn{6}{c}{NGC 4755}\\
  & & & & & \\ 
    H      &     2.634  &  0.295  &  0.014  & -2.49  &  0.54\\
     I-17   &     2.636 & 0.274  & -0.007  & -2.41 &  -0.62\\
     III-06  &    2.639 &   0.269 &  -0.012 &  -2.33 &   0.40\\
     IV-17   &    2.603  &  0.273  & -0.008  & -3.67  & -0.79\\
 & & & &  &\\
\hline
 & & & &  &\\
$mean$& &  & 0.001 & & 0.48\\
$\sigma$  & & & 0.033 & &0.75\\
 & & & &  &\\
\hline
\end{tabular}
\end{center}
\end{table}
%-------------------------------------------------------------------------

%-------------------------------------------------------------------------
\begin{table}
\begin{center}
\caption{Comparison between the mean clusters' reddening and distance
values found in the literature and the mean values obtained with the
present calibration and the FR90 calibration. $DM$ values in columns 5
and 7 do not include the 0.3 mag. correction suggested in Eq. (14).} 
\begin{tabular}{lrrrrrr}
\hline
&&&&&&\\
Cluster & \multicolumn{2}{c}{Literature} & \multicolumn{2}{c}{This work} 
& \multicolumn{2}{c}{FR90} \\
 &{\tiny $E(b-y)$} & {\tiny $DM$} & {\tiny $E(b-y)$} &
{\tiny $DM$} &{\tiny $E(b-y)$} & {\tiny $DM$} \\
&&&&&&\\
\hline
&&&&&&\\
$h$ \& $\chi$ Per & 0.41 &11.4 & 0.42 & 10.9 & 0.41 & 10.4 \\
                  & 2    &  4  &  4   &   5  &    4 &   5  \\
&&&&&&\\
NGC 663 & 0.58 & 12.0 & 0.59 & 11.6 & 0.58 & 10.7 \\
        &  5   &   2  &   4  &   7  &   3  &   3 \\
&&&&&&\\
NGC 3766 & 0.15 & 11.5 & 0.15 & 10.3 &  &  \\
         &  1   &   4  &   5  &  5   &   &  \\
&&&&&&\\
NGC 4755 & 0.28 & 11.4 & 0.28 & 11.5 & & \\
         &  1   &   1  &   1  &   6  & & \\
&&&&&&\\
\hline
\end{tabular}
\end{center}
\end{table} 
%---------------------------------------------------------------------

In column 4 of Table 1 we present the differences between our $E(b-y)$
values and the mean reddening for each cluster. The mean value of the
differences is zero. In Table 2 we present the mean values we obtain for
each cluster. They are the same as the values found in the literature
and determined from observations of the non-emission B stars. We can
conclude that there is not any systematic difference in our calibration.
The
accuracy of our values can be estimated as 0.033 mag., from the deviation
of the differences in Table 1. This value is less than twice the mean
deviation for the $E(b-y)$ determination of normal absorption-line stars,
namely 0.018 mag. (Perry \& Johnston 1982; Franco 1989).

The differences in the determination of the absolute magnitude are shown
in columns 5 and  6 of Table 1. Column 5 presents the absolute
magnitudes obtained from the \uvbyb photometry with our calibration,
which we denote $M_{V}(\beta_{*})$. Column 6 gives the differences
between our determinations and the values obtained from the photometry
and the cluster reddening distance modulus, computed in the following way:

\begin{equation}
\Delta M_{V} = M_{V}(\beta_{*}) - (V_{0} - DM)  
\end{equation}

In this case there is an apparent systematic
difference, our determinations being 0.5 mag. fainter that the values
derived from the distance modulus. The main contribution to this
difference came from stars in NGC 3766. In Table 2 it can be seen that our
distance modulus determination is lower by more than one magnitude than
the value in the literature. This is caused by the difficulty of Equation
(9) to reproduce the unusual main sequence rising verticaly at \co =
0.2, as already stated by Shobbrook (1987). As our luminosity
calibration is based on the assumption of Eq. (9), when this equation is 
not a good representation of the absorption-line stars locus in the
$\beta$ -- \co plane, our calibration is unsuitable. This is the case of
NGC 3766. Hence, within our approach we cannot obtain reliable luminosity
estimations for
stars in this cluster, and consequently we will not consider  it in the
following discussion. It is difficult to know whether this is a
peculiarity of NGC 3766 or if it is common to all clusters
of the same age. More \uvbyb photometry of clusters of similar age 
would be needed to decide on this issue.

If we do not consider stars in NGC 3766, the mean difference of the
absolute magnitudes is 0.32 $\pm$ 0.65. There is still a nonzero mean
difference. However, this value of 0.3 mag. is the expected difference in
\mv between an absorption-line B star and a Be star of the same spectral 
type in the range
B0--B5, as determined by Zorec \& Briot (1991) and already mentioned in
Section 2. Our procedure estimates the absolute magnitude of the
underlying B star without considering the circumstellar contribution. 
The values derived from the uncorrected photometry and the
cluster mean reddening and distance include the circumstellar
contribution. Our values are systematicaly lower by 0.3 mag., confirming
independently the Zorec \& Briot finding that there is an overluminosity
of around this magnitude produced by the circumstellar envelope emission.
Then we can
conclude that there are not systematic effects in our luminosity
calibration. The mean error associated to our \mv determination can be
assumed as 0.7 mag., again lower than twice the mean error of the Balona
\& Shobbrook (1984) \mv calibration for absorption-line B stars, namely
0.43 mag.

The fact that the present calibration gives the \mv value for the
underlying star has to be taken into account when using this value for the
distance determination. In this last case, the mean 0.3 mag.
overluminosity of the Be star has to be considered. The value of the
distance modulus has to be computed in the same way:

\begin{equation}
DM = V_{0} - (M_{V}(\beta_{*}) - 0.3)
\end{equation}

Finally, for comparison we have computed the $E(b-y)$ and \mv values for
stars with simultaneous \uvbyb photometry and \ha spectroscopy using the
preliminary calibration given in FR90. The obtained values are shown in
Tables 2 and 3. For the interstellar reddening the FR90
calibration gives the same results than the present work, with similar
accuracy. The \mv calibration, however, is significantly worse, presenting
a very high systematic error. Even in the case of the interstellar
reddening, the present calibration has the advantage that it is fully
photometric, whereas the FR90 calibration needs the use of simultaneous
\uvby photometry and \ha spectroscopy. Therefore, we consider that the
present work completely supersedes the FR90 calibration.  

%------------------------------------------------------------------------
\begin{table}
\begin{center}
\caption{Comparison between the $E(b-y)$ and \mv values obtained with the
FR90  calibration and the values derived from the clusters' reddening and
distance.} 
\begin{tabular}{rrrrrr}
\hline
  & & & & & \\ 
 Star & $\beta_{*}$ &$E(b-y)$ & \small{$\Delta E(b-y)$} &
$M_{V}(\beta_{*})$ & \small{$\Delta (M_{V})$} \\
  & & & & & \\ 
\hline
&&&&&\\
\multicolumn{6}{c}{$h$ \& $\chi$ Per} \\
  & & & & & \\ 
   309  & 2.639  &  0.391 &  -0.019 &  -2.63  & 0.86\\
   717  &     2.626  &   0.379  &  -0.031  &  -2.77  &   0.87 \\
  1261  &     2.694  &   0.441  &   0.031 &   -1.33 &    2.00\\
  1268  &     2.626   &  0.459 &    0.049  &  -2.83  &   0.71\\
  1702  &     2.636  &   0.381  &  -0.029 &   -2.42 &   0.79\\
  2138 &      2.594 &    0.482 &    0.072 &   -3.74 &   -0.10\\
  2165  &     2.632  &   0.442  &   0.032 &   -2.72  &   0.20\\
  2284  &     2.728 &    0.338  &  -0.072 &   -0.83  &   2.40\\
  2371  &     2.630  &   0.404 &   -0.006 &   -2.68 &    1.01\\
&&&&&\\
\multicolumn{6}{c}{NGC 663}  \\
  & & & & & \\ 
  2    &     2.747  &   0.578  &  -0.012 &   -0.56  &   1.93\\
  6    &     2.773  &   0.611  &   0.021  &  -0.36 &   2.33\\
  10  &      2.686  &   0.613  &   0.023 &   -1.62  &   1.22\\
  21  &      2.630  &   0.588 &   -0.002  &  -2.51 &    1.66\\
  93   &     2.746  &   0.533  &  -0.057 &   -0.38  &   2.39\\
&&&&&\\
\hline
  & & & & & \\ 
$mean$ & & & 0.000 & & 1.30\\
$\sigma$ & & & 0.040 & & 0.82\\
  & & & & & \\ 
\hline
\end{tabular}
\end{center}
\end{table}
%------------------------------------------------------------------------

 \section{Conclusions}
We have used a sample of Be stars in open clusters with simultaneous
\uvbyb photometry and Balmer line spectroscopy to derive
linear relations between the equivalent widths of the Balmer
lines and the anomalies in the \uvby photometric
indices produced by continuum circumstellar emission. Similar relations
exist when the emission in the \hb line is measured through the
photometric $\beta$ index.

These relations have been used to elaborate an empirical
calibration of the \uvbyb photometric system to estimate the relevant
astrophysical parameters of the underlying B
star, valid for spectral types earlier than B5. The proposed calibration
allows the determination of the intrinsic
colours, interstellar reddening and absolute magnitude with estimated
errors lower than twice the errors of the usual \uvbyb calibrations for
absorption-line B stars.

We have also independently confirmed the previous finding that Be stars
have absolute magnitudes brigther by 0.3 mag. in average than the B stars
of the same spectral type. 

\acknowledgements
We wish to express our gratitude to the referee Dr. E.H. Olsen for his
careful reading of the manuscript and valuable comments which improved the
final presentation of the paper.


\begin{thebibliography}{}

   \bibitem{} Balona L.A. 1994, MNRAS 268, 119

   \bibitem{} Balona L.A., Shobbrook R.R. 1984, MNRAS 211, 375

   \bibitem{} Crawford D.L. 1978, AJ 83, 48

   \bibitem{} Crawford D.L., Mander J. 1966, AJ 71, 144

   \bibitem{} Dachs J., Hanuschick R., Kaiser D., Rohe D. 1986, A\&A 
      159, 276

   \bibitem{} Dachs J., Engels D., Kiehling R. 1988, A\&A 194, 167

   \bibitem{} Fabregat J., Reglero V. 1990, MNRAS 247, 407 (FR90)

   \bibitem{} Fabregat J., Torrej\'{o}n J.M. 1997, PASP (submitted)

   \bibitem{} Fabregat J., Torrej\'{o}n J.M., Reig P., et al. 1996, A\&AS
      119, 271 (Paper I)

   \bibitem{} Franco G.A.P. 1989, A\&AS 78, 105

   \bibitem{} Golay M., 1974, Introduction to Astronomical Photometry,
      Reidel, Dordrecht, Holland

   \bibitem{} Kaiser D. 1989, A\&A 222, 187

   \bibitem{} Kurucz R.L. 1979, ApJS 40, 1

   \bibitem{} Perry C.L., Johnston L. 1982, ApJS 50, 451

   \bibitem{} Perry C.L., Franklin Jr. C.B., Landolt A.U. 1976, AJ 81, 632

   \bibitem{} Shobbrook R.R. 1985, MNRAS 212, 591

   \bibitem{} Shobbrook R.R. 1987, MNRAS 225, 999

   \bibitem{} Slettebak A. 1985, ApJS 59, 769

   \bibitem{} Str\"{o}mgren B. 1966, ARAA 4, 433

   \bibitem{} Torrej\'{o}n J.M. 1997, PhD Thesis, University of Valencia.

   \bibitem{} Torrej\'{o}n J.M., Fabregat J., Bernabeu G., Alba S. 1997,
      A\&AS 124, 329 (Paper II)

   \bibitem{} Zorec J., Briot D. 1991, A\&A 245, 150

\end{thebibliography}
\end{document}